\documentclass[a4paper,12pt]{article}
\usepackage{amsmath}
\usepackage{amssymb}
\usepackage{graphicx}
\usepackage{color}
\usepackage{booktabs}
\usepackage{cite}
\usepackage{array,longtable,lscape}

\numberwithin{table}{section}

\begin{document}

\title{Periodic structures described by the perturbed Burgers--Korteweg--de Vries equation}

\author{Nikolai A. Kudryashov, Dmitry I. Sinelshchikov,\\
\texttt{nakudr@gmail.com}, \texttt{disine@gmail.com}, \vspace{0.5cm}\\
Department of Applied Mathematics, \\ National Research Nuclear University MEPhI, \\ 31 Kashirskoe Shosse, 115409 Moscow, Russian Federation}

\date{}

\maketitle

\begin{abstract}
We study the perturbed Burgers--Korteweg--de Vries equation. This equation can be used for the description of nonlinear waves in a liquid with gas bubbles and for the description of nonlinear waves on a fluid layer flowing down an inclined plane. We investigate the integrability of this equation using the Painlev\'{e} approach. We show that the perturbed Burgers--Korteweg--de Vries equation does not belong to the class of integrable equations. Classical and nonclassical symmetries admitted by this equation and corresponding symmetry reductions are constructed. New types of periodic analytical structures described by the Burgers--Korteweg--de Vries equation are found.
\end{abstract}

\noindent
Key words: perturbed Burgers--Korteweg--de Vries equation; nonlinear waves; symmetries; analytical periodic structures.

\section{Introduction}

Nonlinear waves in complex media have been intensively studied during the last decades (see, e.g. \cite{Ablowitz1991,Calogero1991,Porubov,Leblond2008}). There are several prototypical equations that are widely used for the description of nonlinear waves such as the Burgers equation, the Korteweg--de Vries equation and the Burgers--Korteweg--de Vries equation (see, e.g. \cite{Ablowitz1991,Calogero1991,Leblond2008,Polyanin,Saccomandi2014} are references therein). Usually, these 'universal' equations are derived with the help of some asymptotic approach, for instance with the help of the reductive perturbation method (see, e.g. \cite{Frenzen1985,Taniuti1990,Leblond2008}). Taking into account first--order terms in the asymptotic expansion one can obtain famous nonlinear evolution equations like the Kortweg--de Vries equation and the Burgers--Korteweg--de Vries equation. However, if we consider high order corrections in the asymptotic expansion we can obtain generalizations of the above mentioned equations. One can consider these equations as 'universal' equations with high order corrections.

Recently, the Burgers--Korteweg--de Vries equation with high order corrections has been obtained for the description of nonlinear waves in a liquid with gas bubbles \cite{Kudryashov2014}. This equation can be also used for the description of long nonlinear waves on a surface of a fluid layer flowing down an inclined plane \cite{Depassier2012}. The perturbed Burgers--Kortwege--de Vries equation has the form \cite{Kudryashov2014}:
\begin{equation}
\begin{gathered}
v_{t}+ \alpha v v_{x} +\beta v_{xxx}-\mu v_{xx} +\varepsilon \Bigg[(6\beta\alpha-\beta_{2}+3\beta\lambda_{2}+6\beta\lambda_{1}-2\alpha\lambda_{3})v_{x}v_{xx}+\hfill \\
+(2\beta\alpha-\beta_{1}+3\beta\lambda_{2})v v_{xxx}
+\frac{\alpha(2\lambda_{1}+\lambda_{2})}{2}v^{2}v_{x}-(2\mu\lambda_{1}+\mu\alpha+\nu)v_{x}^{2}-\hfill \\-(2\mu\lambda_{2}+\mu\alpha+\nu)vv_{xx}
+\mu^{2} v_{xxx}+(\gamma-3\beta\mu)v_{xxxx}+2\beta^{2} v_{xxxxx}\Bigg]=0.\hfill
\label{eq:extended_BKdV_equation_1}
 \end{gathered}
\end{equation}
Here $v$ is the non--dimensional gas--liquid density perturbation, $t$ is the non--dimensional time, $x$ is the non--dimensional Cartesian coordinate,  $\lambda_{1},\,\lambda_{2},\,\lambda_{3}$ are arbitrary parameters introduced by the near--identity transformations (see \cite{Kudryashov2014}), $\alpha$, $\mu$, $\nu$, $\beta$, $\beta_{1}$, $\beta_{2}$, $\gamma$ are physical parameters, $\varepsilon$ is a small parameter. One can see that if $\epsilon$ goes to zero, Eq. \eqref{eq:extended_BKdV_equation_1} transforms to the Burgers--Korteweg--de Vries equation. Therefore, we will refer to Eq. \eqref{eq:extended_BKdV_equation_1} as a perturbed Burgers--Korteweg--de Vries equation.

Let us briefly discuss some aspects of the derivation of the perturbed Burgers--Korteweg--de Vries equation. As for application to waves in a liquid with gas bubbles, at the derivation of Eq. \eqref{eq:extended_BKdV_equation_1} both high order corrections in the asymptotic expansion and a liquid viscosity, a surface tension, an interphase heat transfer and a weak liquid compressibility have been taken into account \cite{Kudryashov2014}. In the case of a liquid falling down an inclined plane a liquid viscosity and a surface tension have been taken into account along with high order corrections in the asymptotic expansion \cite{Depassier2012}. Thus, we see that the Burgers--Korteweg--de Vries equation with high order corrections can describe nonlinear waves both in dissipative systems and in systems with dissipation and instability. Parameters of Eq. \eqref{eq:extended_BKdV_equation_1} will be discussed below after introducing a canonical form of Eq. \eqref{eq:extended_BKdV_equation_1}.

Note that some particular cases of Eq. \eqref{eq:extended_BKdV_equation_1} have been studied. For instance, neglecting dissipative effects (i.e. setting $\mu=\nu=\gamma=0$) we obtain the generalized Korteweg--de Vries equation \cite{Olver1984,Polyanin} which is used in studying of long waves on shallow water. There are also some other particular cases of Eq. \eqref{eq:extended_BKdV_equation_1} such as the Kawahara equation \cite{Kawahara1972,Polyanin} and the equation for waves in a viscoelastic tube \cite{Kudryashov2011a}. Let us also remark that several solitary wave solutions of Eq. \eqref{eq:extended_BKdV_equation_1} have been found in \cite{Depassier2012}. However, the general case of the perturbed Burgers--Korteweg--de Vries equation has not been investigated thoroughly previously. Therefore, the general case of Eq. \eqref{eq:extended_BKdV_equation_1} is worth studying.

The aim of this work is to study Eq. \eqref{eq:extended_BKdV_equation_1} without imposing conditions on its parameters. To this end we use the Painlev\'{e} approach and the symmetry approach. By means of the Painlev\'{e} approach we show that Eq. \eqref{eq:extended_BKdV_equation_1} is not integrable in the general case. However, applying a truncated Painlev\'{e} expansion we find several classes of solitary waves solutions of Eq. \eqref{eq:extended_BKdV_equation_1}. We consider both classical and nonclassical symmetries of the perturbed Burgers--Korteweg--de Vries equation. We construct corresponding symmetry reductions and their exact solutions. We find that Eq. \eqref{eq:extended_BKdV_equation_1} admits interesting periodic structures which have not been reported previously.

The rest of this work is organized as follows. In Section 2 we transform Eq. \eqref{eq:extended_BKdV_equation_1} into a canonical form and study it using the Painlev\'{e} approach. Section 3 is devoted to the symmetry analysis of the perturbed Burgers--Korteweg--de Vries equation. Both classical and nonclassical symmetries of this equation are studied. In Section 4 we construct symmetry reductions of the perturbed Burgers--Korteweg--de Vries equation and their exact solutions. In the last section we briefly discuss our results.

\section{Canonical form and the Painlev\'{e} test}
\label{sec:2}

In this section we transform Eq. \eqref{eq:extended_BKdV_equation_1} to a canonical form and investigate it using the Painleve test for partial differential equations.

In order to construct a canonical form of Eq. \eqref{eq:extended_BKdV_equation_1} we use the following scaling and shifts transformations:
\begin{equation}
v=A+Bv',\quad x=x'+Et',\quad t=Ft'.
\label{eq:scaling_tranasformations}
\end{equation}
Applying \eqref{eq:scaling_tranasformations} with the following parameters
\begin{equation}
\begin{gathered}
\lambda_{2}=\lambda_{1}, \quad \lambda_{3}=\beta+\frac{1}{2\alpha}(2\beta_{1}+3\beta\lambda_{1}-\beta_{2}), \quad
A=-\frac{1}{3\epsilon\lambda_{1}}, \\
 B=\frac{2}{\alpha\lambda_{1}}(2\beta\alpha-\beta_{1}+3\beta\lambda_{1}), \quad
E=-\frac{\alpha}{12\epsilon^{2}\beta^{2}\lambda_{1}}, \quad F=\frac{1}{2\epsilon\beta^{2}},
 \end{gathered}
\end{equation}
where $\lambda_{1}$ is a solution of the equation
\begin{equation}
(2\beta\alpha-\beta_{1}+3\beta\lambda_{1})^{2}-10\beta^{2}\alpha\lambda_{1}=0,
\end{equation}
from Eq. \eqref{eq:extended_BKdV_equation_1} we obtain the equation (primes are omitted):
\begin{equation}
\begin{gathered}
v_{t}+30 v^{2}v_{x} -\mu_{1} v_{xx}-\sigma v_{xxx}+10 v v_{xxx} +20 v_{x}v_{xx}- \hfill \\
- \mu_{2} (vv_{x})_{x}+\mu_{3} v_{xxxx}+v_{xxxxx}=0.
\label{eq:extended_BKdV_equation_3}
 \end{gathered}
\end{equation}
The parameters $\mu_{1},\mu_{2},\mu_{3}$ and $\sigma$ are given by
\begin{equation}
\begin{gathered}
\mu_{1}=\frac{1}{6\epsilon\beta^{2}\lambda_{1}}(\mu\lambda_{1}-\nu-\mu\alpha),\quad
\mu_{2}=\frac{1}{\beta}\sqrt{\frac{10}{\alpha\lambda_{1}}}(2\mu\lambda_{1}+\mu\alpha+\nu),\\
\mu_{3}=\frac{1}{2\beta^{2}}(\gamma-3\beta\mu),\quad
\sigma=\frac{1}{6\epsilon\beta^{2}\lambda_{1}}(\beta_{1}-2\beta\alpha+3\epsilon \mu^{2}\lambda_{1}).
 \end{gathered}
\end{equation}
Below, we study Eq. \eqref{eq:extended_BKdV_equation_3}.

Let us discuss parameters of Eq. \eqref{eq:extended_BKdV_equation_3}. As far as application of Eq. \eqref{eq:extended_BKdV_equation_3} to waves in a liquid with gas bubbles is concerned, parameters $\mu_{1}$, $\mu_{2}$ and $\mu_{3}$ can be considered as parameters defining dissipation of nonlinear waves governed by Eq. \eqref{eq:extended_BKdV_equation_3}. This dissipation is caused by the inter--phase heat transfer (parameters $\mu_{1}$ and $\mu_{2}$), liquid viscosity (parameters $\mu_{1}$ and $\mu_{2}$) and weak liquid compressibility (the parameter $\mu_{3}$). The parameter $\sigma$ define dispersion of nonlinear waves which is caused by the presence of bubbles and by the interphase heat transfer. There are both dissipation and instability terms in Eq. \eqref{eq:extended_BKdV_equation_3} in the case of long waves on a surface of a liquid flowing down an inclined plane. Parameters $\mu_{1}$ and $\mu_{2}$ define instability caused by gravity and the parameter $\mu_{3}$ defines dissipation caused by the surface tension. Therefore, when application of Eq. \eqref{eq:extended_BKdV_equation_3} to flow of a liquid flowing down an inclined plane is concerned, parameters $\mu_{1}$ and $\mu_{2}$ are negative and the parameter $\mu_{3}$ is positive.

Let us investigate Eq. \eqref{eq:extended_BKdV_equation_3} using the Painlev\'{e} approach for partial differential equations \cite{Weiss1983}. We look for a solution of Eq. \eqref{eq:extended_BKdV_equation_3} in the from \cite{Weiss1983}
\begin{equation}
v=\phi^{p}\sum\limits_{j=0}^{\infty}v_{j}\phi^{j}, \quad \phi\equiv\phi(x,t), \quad v_{j}\equiv v_{j}(x,t).
\label{eq:WTC_expansion}
\end{equation}
The necessary condition for Eq. \eqref{eq:extended_BKdV_equation_3} to possess the Painlev\'{e} property is that expansion \eqref{eq:WTC_expansion} contains five arbitrary functions.

Substituting \eqref{eq:WTC_expansion} into the leading terms of Eq. \eqref{eq:extended_BKdV_equation_3} we find that $p=-2$ and $v_{0}^{(1)}=-2\phi_{x}^{2}$, $v_{0}^{(2)}=-6\phi_{x}^{2}$. Thus, Eq. \eqref{eq:extended_BKdV_equation_3} admits two expansions into series \eqref{eq:WTC_expansion}. Substituting the expression
\begin{equation}
v=v_{0}^{(1,2)}\phi^{-2}+v_{j}\phi^{j-2},
\end{equation}
into the leading terms of Eq. \eqref{eq:extended_BKdV_equation_3} we find the Fuchs indices:
\begin{equation}
\begin{gathered}
j^{(1)}_{1}=-1,\quad j^{(1)}_{2}=2,\quad  j^{(1)}_{3}=5,\quad  j^{(1)}_{4}=6, \quad j^{(1)}_{5}=8,\vspace{0.1cm}\\
j^{(2)}_{1}=-1,\quad j^{(2)}_{2}=-3,\quad  j^{(2)}_{3}=6,\quad  j^{(3)}_{4}=8, \quad j^{(4)}_{5}=10.
 \end{gathered}
\end{equation}
Therefore,  functions $v_{2}, v_{5}, v_{6}$ and $v_{8}$ in the first expansion \eqref{eq:WTC_expansion} have to be arbitrary to Eq. \eqref{eq:extended_BKdV_equation_3} passes the Painlev\'{e} test. However, substituting \eqref{eq:WTC_expansion} into Eq. \eqref{eq:extended_BKdV_equation_3} we obtain that series \eqref{eq:WTC_expansion} exists only when $\sigma=1/84(36\mu_{3}+11\mu_{2})(6\mu_{3}+\mu_{2})$. We also find that functions $v_{2}$ and $v_{5}$ are not arbitrary. Thus, we see that Eq. \eqref{eq:extended_BKdV_equation_3} does not pass the Painlev\'{e} test. In the case of second expansion \eqref{eq:WTC_expansion} there are no additional conditions for series \eqref{eq:WTC_expansion} to exist. However, we find that functions $v_{6}$ and $v_{8}$ are not arbitrary.

\begin{figure}
\centering
\includegraphics[height=5cm,width=6cm]{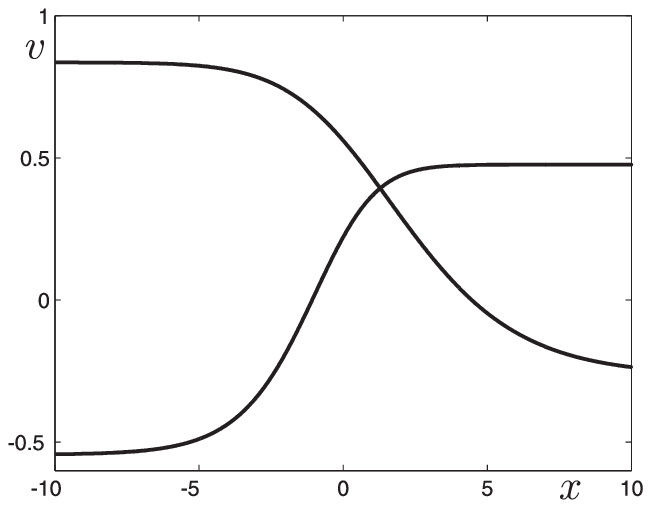}
\includegraphics[height=5cm,width=6cm]{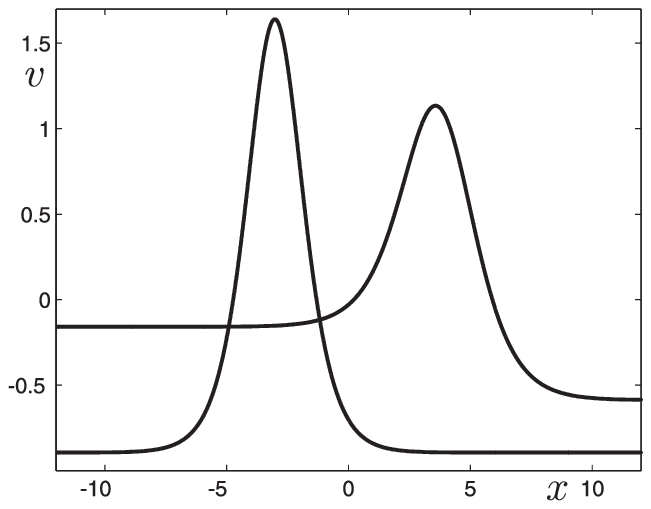}
\caption{Typical kink solutions of Eq.\eqref{eq:extended_BKdV_equation_3} (formulas \eqref{eq:sl_wave_solution_1} and \eqref{eq:sl_wave_solution_2}) (left figure) and typical solitary waves solutions of Eq.\eqref{eq:extended_BKdV_equation_3} (formulas \eqref{eq:sl_wave_solution_4} and \eqref{eq:sl_wave_solution_5}) (right figure).}
\label{f:f1}
\end{figure}

On the other hand, expansion \eqref{eq:WTC_expansion} may be useful for construction of exact solutions of Eq. \eqref{eq:extended_BKdV_equation_3}. We use the following truncation of expansion \eqref{eq:WTC_expansion} \cite{Kudryashov1988,Kudryashov1990,Kudryashov1991,Kudryashov1993}:
\begin{equation}
v=v_{0}^{(1,2)}\phi^{-2}+v_{1}\phi^{-1}+v_{2}.
\label{eq:truncated_expansion}
\end{equation}
Substituting \eqref{eq:truncated_expansion} into Eq. \eqref{eq:extended_BKdV_equation_3} we obtain an overdetermined system of equations for $\phi$, $v_{1}$ and $v_{2}$. In order to find solutions of this system of equations we use the ansatz $\phi=1+\exp\{\theta\}, \theta=k\,x+\omega\,t+\phi_{0}$, where $\phi_{0}$ is an arbitrary constant \cite{Kudryashov1988,Kudryashov1990,Kudryashov1991,Kudryashov1993}. In this way we can find several families of exact solutions of Eq. \eqref{eq:extended_BKdV_equation_3}. For instance, the following typical solitary wave solutions corresponding to both of expansions \eqref{eq:WTC_expansion} can be constructed:
\begin{equation}
\begin{gathered}
v^{(1)}=-\frac{2\phi_{x}^{2}}{\phi^{2}}+\frac{(14k+6\mu_{3}+\mu_{2})\phi_{x}}{7\phi}
+\frac{ ( 444\mu_{3}+95\mu_{2} )  ( 6\mu_{3}+\mu_{2} ) ^{2}-4116\mu_{1}}{1176(75\mu_{3}+16\mu_{2})},
\label{eq:sl_wave_solution_1}
  \end{gathered}
\end{equation}
\begin{equation}
\begin{gathered}
v^{(2)}=-\frac{6\phi_{x}^{2}}{\phi^{2}}+\frac{\sigma}{20}+\frac{(2\mu_{3}+\mu_{2})(368\mu_{3}+79\mu_{2})}{35280},
\label{eq:sl_wave_solution_2}
  \end{gathered}
\end{equation}
\begin{equation}
\begin{gathered}
v^{(3)}=-\frac{6\phi_{x}^{2}}{\phi^{2}}+\frac{(42k-2\mu_{3}-\mu_{2})\phi_{x}}{7\phi}-\frac {11k^{2}}{30}-\\-
\frac {( \mu_{2}+2\mu_{3})( 92\mu_{3}-3780k+25\mu_{2} ) }{52920},
\label{eq:sl_wave_solution_4}
  \end{gathered}
\end{equation}
\begin{equation}
\begin{gathered}
v^{(4)}=-\frac{6\phi_{x}^{2}}{\phi^{2}}+\frac{6k\phi_{x}}{\phi}-\frac{\sigma}{20}-\frac{k^{2}}{2}.
\label{eq:sl_wave_solution_5}
  \end{gathered}
\end{equation}
Values of parameters $k$ and $\omega$ and corresponding constraints on parameters of Eq. \eqref{eq:extended_BKdV_equation_3} are given in Table \ref{t: tab1a}. We present plots of solitary waves solutions described by formulas \eqref{eq:sl_wave_solution_1}--\eqref{eq:sl_wave_solution_5} in Fig.\ref{f:f1}.

Using values of parameters presented in Table \ref{t: tab1a} one can see that there are no constraints on parameters $\mu_{1}$, $\mu_{2}$ and $\mu_{3}$ for
solution \eqref{eq:sl_wave_solution_1} to exist. Thus, a kink--type wave given by \eqref{eq:sl_wave_solution_1} may describe nonlinear waves both in the case of a liquid with gas bubbles and in the case of a fluid flowing down an inclined plane. A balance between nonlinearity, dispersion and dissipation or between nonlinearity, dispersion, dissipation and instability allows the existence of this structure in both of these cases correspondingly. Solution \eqref{eq:sl_wave_solution_2} exists only at $\mu_{1}>0$, and, therefore, may describe kink--type structures in dissipative media such as a liquid with gas bubbles. Solitary wave solutions given by \eqref{eq:sl_wave_solution_4} and \eqref{eq:sl_wave_solution_5}  may appear only in media with dissipation and instability since parameters $\mu_{1}$ and $\mu_{2}$ are negative. Therefore, only a balance between nonlinearity, dispersion, dissipation and instability leads to existence of solitary wave solutions. 

\section{Symmetry analysis}

In this section we study symmetries admitted by Eq. \eqref{eq:extended_BKdV_equation_3}. To this end we use the classical Lie method (see, e.g. \cite{Olver,Ibragimov}) and the nonclassical method by Bluman and Cole \cite{Bluman1969}.

To study symmetries of Eq. \eqref{eq:extended_BKdV_equation_3} it is convenient to transform this equation into the potential form. Using the variable $v=w_{x}$ from Eq. \eqref{eq:extended_BKdV_equation_3} we obtain
\begin{equation}
\begin{gathered}
\Delta(w)=w_{t}+10w_{x}^{3}-\mu_{1} w_{xx}-\sigma w_{xxx}+10 w_{x}w_{xxx}+5w_{xx}^{2}-\\-\mu_{2} w_{x}w_{xx}+\mu_{3} w_{xxxx}+w_{xxxxx}=0.
\end{gathered}
\label{eq:extended_BKdV_equation_p}
\end{equation}
Below, we consider symmetries of Eq. \eqref{eq:extended_BKdV_equation_p}.

\subsection{Classical symmetries}

Let us apply the classical Lie method to Eq. \eqref{eq:extended_BKdV_equation_p}. It is known that Eq. \eqref{eq:extended_BKdV_equation_p} is invariant under action of the infinitesimal transformations
\begin{equation}
\tilde{x}=x+a\xi(x,t,w), \quad \tilde{t}=t+a\tau(x,t,w), \quad \tilde{w}=w+a\eta(x,t,w),
\end{equation}
where $a$ is the group parameter, if the determining equations are satisfied on the solutions of Eq. \eqref{eq:extended_BKdV_equation_p}
\begin{equation}
X^{(5)}\Delta\big|_{\Delta=0}=0.
\end{equation}
Here $X^{(5)}$ is the prolonged infinitesimal generator \cite{Olver,Ibragimov}.

Substituting the expression for $X^{(5)}$ we obtain an overdetermined system of linear partial differential equations for infinitesimals $\xi$, $\tau$ and $\eta$. Solving this system of equations we find that Eq. \eqref{eq:extended_BKdV_equation_p} admits three one-parametric Lie groups with the following infinitesimal generators:
\begin{equation}
X_{1}=\partial_{x}, \quad X_{2}=\partial_{t}, \quad X_{3}=\partial_{w}.
\end{equation}
In the case of $\mu_{1}=\mu_{2}=\mu_{3}=\sigma=0$ we find additional infinitesimal generators:
\begin{equation}
X_{4}=x\partial_{x}, \quad X_{5}=5t\partial_{t}, \quad X_{6}=-w\partial_{w}.
\end{equation}
Infinitesimal generators $X_{1}$, $X_{2}$ and $X_{3}$ correspond to shifts in $x$, $t$ and $w$ correspondingly, while infinitesimal generators $X_{4}$, $X_{5}$ and $X_{6}$ correspond to invariance of Eq. \eqref{eq:extended_BKdV_equation_p} under scaling transformations. Therefore, Eq. \eqref{eq:extended_BKdV_equation_p}, and consequently Eq. \eqref{eq:extended_BKdV_equation_3}, admits only the traveling wave reduction in the general case.  The case of $\mu_{1}=\mu_{2}=\mu_{3}=\sigma=0$ corresponds to the self--similar reduction of the generalized Kortweg--de Vries equation.

\subsection{Nonclassical symmetries}

It is known that partial differential equations may admit symmetries which cannot be found with the classical Lie approach \cite{Bluman1969,Popovych1999,Saccomandi2000,Nucci2003,Saccomandi2004,Gandarias2012}. Such symmetries are called nonclassical symmetries. Below, we investigate these symmetries using the method by Bluman and Cole \cite{Bluman1969}.

According to this method we consider the additional auxiliary equation
\begin{equation}
\xi w_{x}+\tau w_{t}-\eta=0.
\label{eq:BC_1}
\end{equation}
Eq. \eqref{eq:BC_1} is the invariant surface condition associated with the vector field $X=\xi \partial_{x}+\tau \partial_{t}+\eta \partial_{w}$. Then we look for classical symmetries admitted by Eqs. \eqref{eq:extended_BKdV_equation_p} and \eqref{eq:BC_1} simultaneously.

\setlength{\extrarowheight}{5pt}
\begin{longtable}{|c|p{4cm}|p{9cm}|}
\caption{Nonclassical infinitesimals admitted by Eq. \eqref{eq:extended_BKdV_equation_3}.} \label{t: tab1} \\ \hline
\textbf{n} & Parameters & $\eta$   \\ \hline \endfirsthead   \hline
\textbf{n} & Parameters & $\eta$   \\  \hline \endhead
1 &\ $\mu_{2}=-6\mu_{3},\,\,\, \sigma=0$ & $\eta^{1}=-\frac{w^{2}}{2}+c_{1}w+\frac{\mu_{1}}{2\mu_{3}}-\frac{c_{1}^{2}}{2}$
 \\ \hline
2 &\ $\mu_{2}=-2\mu_{3}$,\, $\sigma=-\frac{10\mu_{1}}{\mu_{3}}$& $\eta^{2}=-\frac{w^{2}}{6}+c_{2}w-\frac{3c_{2}^{2}}{2}+\frac{3\mu_{1}}{2\mu_{3}}$
 \\ \hline
3 &\ $\mu_{2}=-2\mu_{3}$,\, $\sigma-\frac{10\mu_{1}}{\mu_{3}}$  & $\eta^{3}=-\frac{w^{2}}{6}+(c_{3}x-90c_{3}^{3}t)w-\frac{3c_{3}^{2}}{2}(x-90c_{3}^{2}t)^{2}-\frac{3}{2\mu_{3}}(4\mu_{3}c_{3}-\mu_{1})$ \\  \hline
4 &\ $\mu_{2}=-6\mu_{3}$,\, $\sigma=0$ & $\eta^{4}=-2\left(x-\frac{5\mu_{1}^{2}}{6\mu_{3}^{2}}t+c_{4}\right)^{-2}+\frac{\mu_{1}}{6\mu_{3}}$
 \\ \hline
5 &\ $\mu_{2}=-2\mu_{3}$,\, $\mu_{1}=-\frac{\sigma\mu_{3}}{10}$ & $\eta^{5}=-6\left(x-\frac{3\sigma^{2}}{40}t+c_{5}\right)^{-2}-\frac{\sigma}{20}$
 \\ \hline
\end{longtable}

It is known that if $X$ is a nonclassical symmetry generator then $\lambda X$ is a nonclassical symmetry generator as well for any function $\lambda(x,t,w)\neq 0$ (see, e.g. \cite{Popovych1999,Popovych2008}). Thus, further we have to consider two cases of nonclassical symmetry generators. The first one is the case of $\tau\neq0$, where without loss of generality we can assume that $\tau=1$. The other one is the case of $\tau=0$, where without loss of generality we can assume that $\xi=1$. The case of  $\tau\neq0$ is called the regular case where we have an overdetermined system of equations for infinitesimals. In the singular case $\tau=0$ we obtain a single partial differential equation for infinitesimal $\eta$. As it was shown in \cite{Popovych2008} every solution of this equation generates a family of solutions of a considered equation.

Let us consider the case of $\tau\neq0$. Assuming that $\tau=1$ and applying the classical Lie method to Eqs. \eqref{eq:extended_BKdV_equation_p} and \eqref{eq:BC_1} we obtain an overdetermined system of nonlinear partial differential equations for infinitesimals $\xi$ and $\eta$. Solving this system of equations we find that corresponding infinitesimals give us classical symmetries of Eq. \eqref{eq:extended_BKdV_equation_p}.

Now we study the case of $\tau=1$. Without loss of generality we assume that $\xi=1$ and applying the classical Lie method to Eqs. \eqref{eq:extended_BKdV_equation_p} and \eqref{eq:BC_1} we get a nonlinear partial differential equation for infinitesimal $\eta$. It seems impossible to find the general solution of this equation, however every particular solution of this equation generates nonclassical symmetry reduction of Eq. \eqref{eq:extended_BKdV_equation_p} and corresponding one--parametric family of solutions.

We look for some particular solutions of the equation for $\eta$ and present our results in Table \ref{t: tab1}, where $c_{1},\, i=1,\ldots 5$ are arbitrary constants.  Symmetry reductions corresponding to infinitesimals $\eta^{i}$, $i=1,\ldots5$ will be discussed in the next section.

\section{Symmetry reductions and exact solutions}

In this section we consider traveling wave reduction of Eq. \eqref{eq:extended_BKdV_equation_3}. We also discuss reductions corresponding to nonclassical symmetries admitted by Eq. \eqref{eq:extended_BKdV_equation_3}.

\subsection{Traveling wave solutions}

Let us construct traveling wave solutions of Eq. \eqref{eq:extended_BKdV_equation_3}. Using the variables $v(x,t)=y(z),\,z=x-C_{0}t$ in \eqref{eq:extended_BKdV_equation_3} and integrating the result with respect to $z$ we get
\begin{equation}
C_{1}-C_{0}y+10y^{3}-\mu_{1}y_{z}-\mu_{2} y y_{z}-\sigma y_{zz}+10y y_{zz}+5 y_{z}^{2}+ \mu_{3} y_{zzz}+y_{zzzz}=0,
\label{eq:ExtBKdV_tw}
\end{equation}
where $C_{1}$ is an integration constant.

Note that traveling wave solutions in the form of solitary waves have been constructed in Section \ref{sec:2} and in \cite{Depassier2012}. Here we construct elliptic solutions of Eq. \eqref{eq:ExtBKdV_tw}. To this end we use an approach proposed in \cite{Kudryashov2010,Kudryashov2011,Kudryashov2012,Kudryashov2014b,Kudryashov2014a}. Let us briefly describe this approach. At the first step we construct Laurent expansion for a solution of Eq. \eqref{eq:ExtBKdV_tw} in a neighborhood of movable singular points. Then, using this information we choose corresponding form of a possible elliptic solution with arbitrary parameters. At the last step, we find values of these parameters solving an algebraic system of equations for these parameters. One can obtain this system of equations by expanding possible elliptic solution into Laurent series, substituting  the result into Eq. \eqref{eq:ExtBKdV_tw} and equating coefficients at different powers of $z$ to zero.

Eq. \eqref{eq:ExtBKdV_tw} admits two different expansions into Laurent series, which are the following
\begin{equation}
y^{(1)}=-\frac{2}{z^{2}}+\frac{\mu_{2}+6\mu_{3}}{7z}+\ldots,\quad
y^{(2)}=-\frac{6}{z^{2}}-\frac{\mu_{2}+2\mu_{3}}{7z}+\ldots\ .
\label{eq:tw_expnasion}
\end{equation}
The necessary condition for elliptic solutions to exist is that the sum of residues of solutions \eqref{eq:tw_expnasion} in a neighborhood of poles is zero. Therefore, elliptic solutions of Eq. \eqref{eq:ExtBKdV_tw} may exist in the following cases: the case of $\mu_{2}=-6\mu_{3}$ which corresponds to the first expansion \eqref{eq:tw_expnasion}, the case of $\mu_{2}=-2\mu_{3}$ which corresponds to the second expansion \eqref{eq:tw_expnasion} and the case of $\mu_{3}=0$ which corresponds to both expansions \eqref{eq:tw_expnasion}.

\begin{figure}
\centering
\includegraphics[]{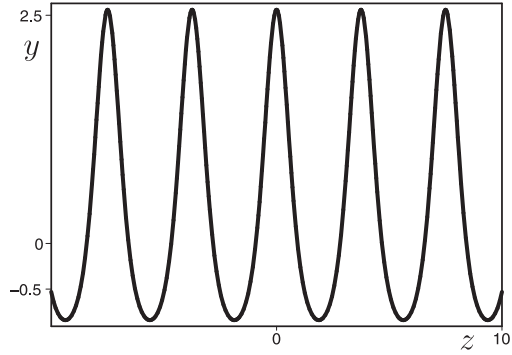}
\caption{Second order elliptic solution \eqref{eq:tw_case_1_so_elliptic} of Eq.\eqref{eq:extended_BKdV_equation_3}.}
\label{f:f2}
\end{figure}

Note that the order of an elliptic function is the number of poles in a parallelogram of periods. Since expansions \eqref{eq:tw_expnasion} contain arbitrary constants Eq. \eqref{eq:ExtBKdV_tw} may admit an elliptic solution of any order. Here we consider second--order, fourth--order and six--order elliptic solutions of Eq. \eqref{eq:ExtBKdV_tw}.

\begin{figure}
\centering
\includegraphics[]{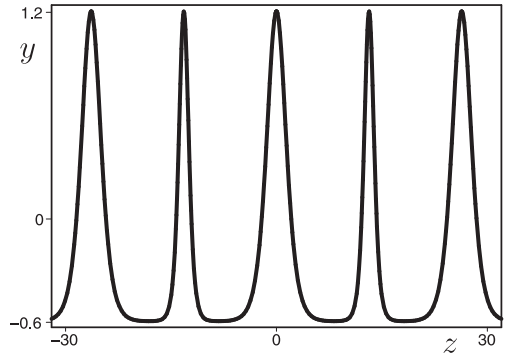}
\caption{Fourth order elliptic solution \eqref{eq:tw_case_2_fo_elliptic} of Eq.\eqref{eq:extended_BKdV_equation_3}.}
\label{f:f3}
\end{figure}

Let us study the case of $\mu_{2}=-6\mu_{3}$. Second, fourth and six order elliptic solutions of Eq. \eqref{eq:ExtBKdV_tw} corresponding to expansion $y^{(1)}$ have the form
\begin{equation}
\begin{gathered}
y^{(1,1)}=-2\wp+\frac{\mu_{1}}{6\mu_{3}},
 \label{eq:tw_case_1_so_elliptic}
 \end{gathered}
\end{equation}
\begin{equation}
\begin{gathered}
y^{(1,2)}=-\frac{1}{2}\left[\frac{\wp_{z}}{\wp-A}\right]^{2}+4A+\frac{\mu_{1}}{6\mu_{3}},
 \label{eq:tw_case_1_fo_elliptic}
 \end{gathered}
\end{equation}
\begin{equation}
\begin{gathered}
y^{(1,2)}=-\frac{1}{2}\left(\left[\frac{\wp_{z}+B}{\wp-A}\right]^{2}+\left[\frac{\wp_{z}-B}{\wp-A}\right]^{2}\right)+2\wp+8A+\frac{\mu_{1}}{6\mu_{3}},
 \label{eq:tw_case_1_six_o_elliptic}
 \end{gathered}
\end{equation}
Note that we use the following notation $\wp\equiv\wp\{z-z_{0},g_{2},g_{3}\}$ throughout this work, where $z_{0}$ is an arbitrary constant. The values of $g_{2}$, $g_{3}$ and corresponding constraints on the parameters of Eq. \eqref{eq:extended_BKdV_equation_3} are given in Table \ref{t: tab3}. Here and below $A$ and $B$ are arbitrary constants.

Now we proceed to the case of $\mu_{2}=-2\mu_{3}$. We find the following second, fourth and sixth order elliptic solutions of Eq. \eqref{eq:ExtBKdV_tw}:
\begin{equation}
\begin{gathered}
y^{(2,1)}=-6\wp+\frac{\mu_{1}}{2\mu_{3}},
  \label{eq:tw_case_2_so_elliptic}
 \end{gathered}
\end{equation}
\begin{equation}
\begin{gathered}
y^{(2,2)}=-\frac{3}{2}\left[\frac{\wp_{z}}{\wp-A}\right]^{2}+12A+\frac{\mu_{1}}{2\mu_{3}},
  \label{eq:tw_case_2_fo_elliptic}
 \end{gathered}
\end{equation}
\begin{equation}
\begin{gathered}
y^{(2,3)}=-\frac{3}{2}\left(\left[\frac{\wp_{z}+B}{\wp-A}\right]^{2}+\left[\frac{\wp_{z}-B}{\wp+A}\right]^{2}\right)+6\wp+\frac{\mu_{1}}{2\mu_{3}}+24A,
  \label{eq:tw_case_2_six_o_elliptic}
 \end{gathered}
\end{equation}

\begin{figure}
\centering
\includegraphics[]{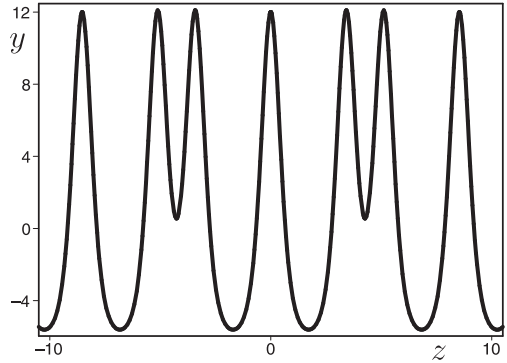}
\caption{Six order elliptic solution \eqref{eq:tw_case_2_six_o_elliptic} of Eq.\eqref{eq:extended_BKdV_equation_3}.}
\label{f:f4}
\end{figure}

In the case of $\mu_{3}=0$ we have the following fourth--order elliptic solution corresponding to both expansions \eqref{eq:tw_expnasion}
\begin{equation}
\begin{gathered}
y^{(1,2)}=-\frac{3}{2}\left[\frac{\wp_{z}+B}{\wp-A}\right]^{2}-\frac{\mu_{2}}{14}\frac{\wp_{z}+B}{\wp-A}+4\wp-\frac{451133\mu_{2}^2}{16458120},
  \label{eq:tw_case_3_fo_elliptic}
 \end{gathered}
\end{equation}
Parameters $g_{2}$, $g_{3}$ of solutions \eqref{eq:tw_case_2_so_elliptic}--\eqref{eq:tw_case_3_fo_elliptic} and corresponding constraints on parameters of Eq. \eqref{eq:extended_BKdV_equation_3} are given in Table \ref{t: tab3}.

Let us discuss elliptic solutions which have been obtained above. Taking into account values of parameters from Table \ref{t: tab3} one can see that all elliptic solutions obtained above exist only when $\mu_{2}<0$ or $\mu_{1}<0$. Thus, periodic structures described by Eq. \eqref{eq:extended_BKdV_equation_3} may emerge only in media with dissipation and instability, for example on a surface of a liquid flowing down an inclined plane. A balance between dissipation, instability, dispersion and nonlinearity leads to emerging of such structures. It is also worth noting that these periodic structures do not appear in purely dissipative media such as a liquid with gas bubbles.

Typical profiles of periodic structures given by \eqref{eq:tw_case_1_so_elliptic}--\eqref{eq:tw_case_3_fo_elliptic} are shown in Figs. \ref{f:f2}--\ref{f:f4}. One can see that second order elliptic solutions describe usual periodic solution, while fourth order and sixth order elliptic solutions describe more complicated periodic structures. In Fig. \ref{f:f3} we see periodic structure with peaks having two different widths. In Fig. \ref{f:f4} we see periodic structure which represents periodic two--crests wave. Thus, we see that in media with dissipation and instability may exist periodic waves with complicated structure. We believe that these periodic structures are reported for the first time. 

\subsection{Nonclassical reductions}

Let us consider nonclassical reductions corresponding to infinitesimals presented in Table \ref{t: tab1}. Infinitesimals $\eta^{4}$ and $\eta^{5}$ correspond to simple rational solutions of Eqs. \eqref{eq:extended_BKdV_equation_p} and \eqref{eq:extended_BKdV_equation_3} and we do not consider associated symmetry reductions. Let us consider infinitesimals $\eta^{1}$, $\eta^{2}$ and $\eta^{3}$.

In the case of $\eta^{1}$ similarity variable for Eq. \eqref{eq:extended_BKdV_equation_p} at $\mu_{2}=-6\mu_{3},\, \sigma=0$ have the form
\begin{equation}
w=c_{1}+\sqrt{\frac{\mu_{1}}{\mu_{3}}}\tanh\left\{\frac{1}{2}\sqrt{\frac{\mu_{1}}{\mu_{3}}}(x+h(t))\right\},
\end{equation}
where function $h(t)$ satisfies the equation
\begin{equation}
h_{t}+\frac{\mu_{1}^{2}}{\mu_{3}^{2}}=0.
\end{equation}
Solving this equation for $h(t)$ we find the following solution of Eq. \eqref{eq:extended_BKdV_equation_p}
\begin{equation}
w=c_{1}+\sqrt{\frac{\mu_{1}}{\mu_{3}}}\tanh\left\{\frac{1}{2}\sqrt{\frac{\mu_{1}}{\mu_{3}}}\left(x-\frac{\mu_{1}^{2}}{\mu_{3}^{2}}t+t_{0}\right)\right\},
\end{equation}
where $t_{0}$ is an arbitrary constant. Using the relation $u=w_{x}$ we obtain solitary wave solution of \eqref{eq:extended_BKdV_equation_3} in the form
\begin{equation}
v=\frac{\mu_{1}}{2\mu_{3}}\left(1-\tanh^{2}\left\{\frac{1}{2}\sqrt{\frac{\mu_{1}}{\mu_{3}}}\left(x-\frac{\mu_{1}^{2}}{\mu_{3}^{2}}t+t_{0}\right)\right\}\right).
\label{eq:nc_sl_sol}
\end{equation}

In the same way we find the following exact solutions of Eq. \eqref{eq:extended_BKdV_equation_3} corresponding to infinitesimals $\eta^{2}$ and $\eta^{3}$:
\begin{equation}
v=\frac{3\mu_{1}}{2\mu_{3}}\left(1-\tanh^{2}\left\{\frac{1}{2}\sqrt{\frac{\mu_{1}}{\mu_{3}}}\left(x-\frac{11\mu_{1}^{2}}{\mu_{3}^{2}}t+t_{0}\right)\right\}\right),
\label{eq:nc_sl_sol1}
\end{equation}
\begin{equation}
v=\frac{3}{2\mu_{3}}\left[(6\mu_{3}c_{3}-\mu_{1})\tan^{2}\left\{x-\frac{126\mu_{3}^{2}c_{3}^{2}-42\mu_{3}\mu_{1}c_{3}+11\mu_{1}^{2}}{\mu_{3}^{2}}t+t_{0}\right\}
-2\mu_{3}c_{3}\right].
\label{eq:nc_sl_sol3}
\end{equation}
Thus, one can see that nonclassical reductions of Eq. \eqref{eq:extended_BKdV_equation_3} corresponding to infinitesimals $\eta^{i}, i=1,2,3$ generate traveling wave solutions, which have been studied above.

In this section we have studied symmetry reductions of Eq. \eqref{eq:extended_BKdV_equation_3}. Elliptic traveling wave solution of Eq. \eqref{eq:extended_BKdV_equation_3} have been constructed. Some nonclassical reductions of Eq. \eqref{eq:extended_BKdV_equation_3} have been considered.

\section{Conclusion}
We have analytically studied the perturbed Burgers--Korteweg--de Vries equation. Using the Painlev\'{e} approach we have shown that this equation does not belong to the class of integrable equations. We have constructed some solitary wave solutions using the truncation procedure in the Painlev\'{e} approach.  We have shown that kink--type structures described by \eqref{eq:extended_BKdV_equation_3} may exist both in media with dissipation and in media with dissipation and instability, while solitary wave structures exist only in media with dissipation and instability.   We have studied classical and nonclassical symmetries admitted by the perturbed Burgers--Korteweg--de Vries equation. It has been shown that in the general case this equation admits only simple shift transformations in $x$ and $t$. Several nonclassical symmetries of the perturbed Burgers--Korteweg--de Vries equation have been found as well. We have constructed some classes of elliptic traveling wave solutions of the perturbed Burgers--Korteweg--de Vries equation. We have found that these solutions describe new periodic structures.  We have demonstrated that obtained periodic structures may exist only in media with dissipation and instability.

\section{Acknowledgments}
This research was partially supported by grant for Scientific Schools 2296.2014.1., by
grant for the state support of young Russian scientists 3694.2014.1 and by RFBR grants 14--01--00498, 14--01--31078.

\appendix

\section{Parameters of solitary wave solutions}

Parameters of solutions \eqref{eq:sl_wave_solution_1}--\eqref{eq:sl_wave_solution_5} and corresponding constraints on parameters of Eq. \eqref{eq:extended_BKdV_equation_3} are given in Table \ref{t: tab1a}.

\setlength{\extrarowheight}{5pt}
\begin{longtable}{|p{1.1cm}|p{14.7cm}|}
\caption{Solitary wave solution parameters} \label{t: tab1a} \\ \hline
Solution & Parameters   \\ \hline \endfirsthead   \hline
Solution & Parameters    \\  \hline \endhead
\eqref{eq:sl_wave_solution_1} & \begin{tabular}[l]{l} $\sigma=\frac{1}{84}(36\mu_{3}+11\mu_{2})(6\mu_{3}+\mu_{2})$,\ \ \  $k=-\frac{1}{14}(6\mu_{3}+\mu_{2})$,\ \ \
$\omega=\frac {5 ( 6\mu_{3}+\mu_{2} ) } { 460992( 75\mu_{3}+16\mu_{2}) ^{2}}\times$ \\$ \times
\left(2420208\mu_{1}^{2}+1176( 6\mu_{3}+\mu_{2} )^{3}\mu_{1}+( 9648\mu_{3}^{2}+4116\mu_{3}\mu_{2}+439\mu_{2}^{2} )( 6\mu_{3}+\mu_{2} ) ^{4} \right)$
\end{tabular} \\ \hline
\eqref{eq:sl_wave_solution_2} & \begin{tabular}[l]{l} $k=\frac{1}{42}(\mu_{2}+2\mu_{3})$,\ \ \ $\mu_{1}= \frac {2\mu_{3}+\mu_{2}}{17640}
 \left( 23\mu_{2}^{2}+381\mu_{3}\mu_{2}+880\mu_{3}^{2} \right)- \frac{\sigma ( 50\mu_{3}+18\mu_{2})}{140}$, \\ $
\omega=-\frac { \left( 2\mu_{3}+\mu_{2} \right)}{248935680}  \big( 444528
\sigma^{2}-504 ( 2\mu_{3}+\mu_{2} )  ( 248\mu_{3}+19\mu_{2}) \sigma+$ \\ $+( 9472\mu_{3}^{2}+2032\mu_{3}\mu_{2}+223
\mu_{2}^{2} )  ( 2\mu_{3}+\mu_{2} ) ^{2}\big)$\end{tabular} \\ \hline
\eqref{eq:sl_wave_solution_4} & \begin{tabular}[l]{l}
$\sigma=-\frac{8k^{2}}{3}+\frac {( \mu_{2}+2\mu_{3} )}{5292}  (988\mu_{3}+137\mu_{2} )$,\ \ \
$\mu_{1}=-\frac{k^{2}( 150\mu_{3}+103\mu_{2} )}{210}-$\\ $-\frac {(23\mu_{2}+60\mu_{3})( 25\mu_{2}+92
\mu_{3} )  ( \mu_{2}+2\mu_{3} ) }{370440}$,\ \ \ 
$\omega=-\frac {121k^{5}}{30}-\frac {k ( \mu_{2}+2
\mu_{3} ) }{93350880} (  3528( 3712\mu_{3}+1625\mu_{2}) k^{2}$\\$+
( \mu_{2}+2\mu_{3} )  ( 25\mu_{2}+92\mu_{3}) ^{2} )$
\end{tabular} \\ \hline
\eqref{eq:sl_wave_solution_5} & \begin{tabular}[l]{l}
$\mu_{2}=-2\mu_{3}$,\,\,\,$\mu_{1}=-\frac{\sigma\mu_{3}}{10}$,\,\,\, $\omega=-\frac{k}{40}(3\sigma^{2}+140k^{4})$
\end{tabular} \\ \hline
\end{longtable}

\section{Parameters of elliptic solutions}

Parameters of solutions \eqref{eq:tw_case_1_so_elliptic}--\eqref{eq:tw_case_3_fo_elliptic} and corresponding constraints on parameters of Eq. \eqref{eq:extended_BKdV_equation_3} are given in Table \ref{t: tab3}.

\setlength{\extrarowheight}{5pt}
\begin{longtable}{|c|p{12.5cm}|}
\caption{Parameters of elliptic solutions of Eq. \eqref{eq:ExtBKdV_tw}} \label{t: tab3} \\ \hline
\textbf{Solution} & \textbf{Parameters}  \\ \hline \endfirsthead   \hline
\textbf{Solution} & \textbf{Parameters} \\  \hline \endhead
\multicolumn{2}{|c|} {The case of $\mu_{2}=-6\mu_{3}$.}  \\  \hline
\eqref{eq:tw_case_1_so_elliptic} &\ $g_{2}=\frac{1}{12\mu_{3}^{2}}(6C_{0}\mu_{3}^{2}-5\mu_{1}^{2}),\,\,\,
 g_{3}=\frac{\mu_{1}}{216\mu_{3}^{3}}(35\mu_{1}^{2}-36C_{0}\mu_{3}^{2})-\frac{C_{1}}{4},\,\,\, \sigma=0$ \\ \hline
\eqref{eq:tw_case_1_fo_elliptic} &\ \begin{tabular}[l]{l} $\sigma=0,\,\,\,
g_{2}=\frac{1}{48\mu_{3}^{2}}(5\mu_{1}^{2}-6C_{0}\mu_{3}^{2})+15A^{2},$ $
g_{3}=\frac{A}{48\mu_{3}^{2}}(6C_{0}\mu_{3}^{2}-528A^{2}\mu_{3}^{2}-$\\ $-5\mu_{1}^{2})$,
$C_{1}=128A^{3}-\frac{2C_{0}}{3\mu_{3}}(\mu_{1}+6A\mu_{3})+\frac{5\mu_{1}^{2}}{54\mu_{3}^{3}}(7\mu_{1}+36A\mu_{3})$ \end{tabular} \\ \hline
\eqref{eq:tw_case_1_six_o_elliptic}& \begin{tabular}[l]{l} $\sigma=0, \,\,\, g_{2}=12A^{2}-4B\sqrt{3A},\,\,\, g_{3}=-8A^{3}-B^{2}+4B\sqrt{3A^{3}}$,\\$
C_{0}=24A^{2}+72B\sqrt{3A}+\frac{5\mu_{1}^{2}}{6\mu_{3}^{2}}$, \,\,\,
$C_{1}=32A^{3}-108B^{2}-240B\sqrt{3A^{3}}-$\\$-\frac{\mu_{1}}{54\mu_{3}^{3}}(864\mu_{3}^2A^2+2592 B \mu_{3}^2 \sqrt{3A}-5\mu_{1}^2)$ \end{tabular}  \\ \hline
\multicolumn{2}{|c|} {The case of $\mu_{2}=-2\mu_{3}$.} \\ \hline
\eqref{eq:tw_case_2_so_elliptic} &  $\sigma=-\frac{10\mu_{1}}{\mu_{3}},\,\,\, g_{2}=\frac{1}{84\mu_{3}^{2}}(2C_{0}\mu_{3}^{2}-15\mu_{1}^{2}), \,\,\,
 g_{3}=\frac{\mu_{1}}{1512\mu_{3}^{3}}(8C_{0}\mu_{3}^{2}-95\mu_{1}^{2})+\frac{C_{1}}{108}$  \\ \hline
\eqref{eq:tw_case_2_fo_elliptic} & \begin{tabular}[l]{l} $\sigma=-\frac{10\mu_{1}}{\mu_{3}}, \,\,\,
g_{2}=15A^{2}+\frac{5\mu_{1}^{2}}{112\mu_{3}^{2}}-\frac{C_{0}}{168}, \,\,\,
g_{3}=\frac{C_{0}A}{168}-11A^{3}-\frac{5A\mu_{1}^{2}}{112\mu_{3}^{2}}$,\\$
C_{1}=\frac{4}{7\mu_{3}}(9\mu_{3}A-\mu_{1})C_{0}-3456A^{3}+\frac{5\mu_{1}^{2}}{14\mu_{3}^{3}}(19\mu_{1}-108\mu_{3}A)$  \end{tabular}  \\ \hline
\eqref{eq:tw_case_2_six_o_elliptic} & \begin{tabular}[l]{l} $g_{2}=12A^{2}-4\sqrt{3A}B, \,\,\, g_{3}=4\sqrt{3A^{3}}B-8A^{3}-B^{2},\,\,\, \sigma=-\frac{10\mu_{1}}{\mu_{3}},$\\$
C_{0}=\frac{15\mu_{1}^{2}}{2\mu_{3}^{2}}+504 A^{2}+1512\sqrt{3A}B$,\,\,\, $C_{1}=\frac{5\mu_{1}^{3}}{2\mu_{3}^{3}}-864A^{3}-$\\$-\frac{288\mu_{1}}{\mu_{3}}(A^{2}+3B\sqrt{3A})+324B(9B+20\sqrt{3A^{3}})$ \end{tabular} \\ \hline
\multicolumn{2}{|c|} {The case of $\mu_{3}=0$.}  \\  \hline
\eqref{eq:tw_case_3_fo_elliptic} & \begin{tabular}[l]{l} $A=-\frac{17845\mu_{2}^2}{6583248},\,\,\, B=\frac{199\mu_{2}^{3}}{1097208},\,\,\,
\sigma=\frac{11\mu_{2}^{2}}{84},\,\,\, \mu_{1}=-\frac{36607\mu_{2}^{3}}{2743020},$ \\
$g_{2}=\frac{205677481\mu_{2}^{4}}{3611596185792}$, $g_{3}=\frac{1490990741389\mu_{2}^{6}}{35664050050384218624},\,\,\,
C_{0}=\frac{9373459\mu_{2}^{4}}{1279618830}$,\\$ C_{1}=-\frac{2269740279082\mu_{2}^{6}}{64496618291348775}$ \end{tabular}  \\ \hline
\end{longtable}

\end{document}